\documentclass[prb,twocolumn,notitlepage,showpacs,amsmath,amstex,amssymb,citeautoscript,longbibliography,superscriptaddress,
]{revtex4-1}
\pdfoutput=1

\usepackage{natbib}
\usepackage[english]{babel}
\usepackage{letltxmacro}
\usepackage{latexsym}
\LetLtxMacro{\ORIGselectlanguage}{\selectlanguage}
\makeatletter
\DeclareRobustCommand{\selectlanguage}[1]{
  \@ifundefined{alias@\string#1}
    {\ORIGselectlanguage{#1}}
    {\begingroup\edef\x{\endgroup
       \noexpand\ORIGselectlanguage{\@nameuse{alias@#1}}}\x}%
}
\newcommand{\definelanguagealias}[2]{%
  \@namedef{alias@#1}{#2}%
}
\makeatother
\definelanguagealias{en}{english}
\definelanguagealias{English}{english}
\usepackage{graphicx}
\usepackage{amsmath}
\usepackage{mathtools}
\usepackage{amsfonts}
\usepackage{amssymb,bbding}
\usepackage{bm}
\usepackage{color}
\usepackage[percent]{overpic}
\usepackage{soul} 
\usepackage{amssymb}
\usepackage{wasysym}
\usepackage{dsfont}
\usepackage{float}
\usepackage{braket}
\usepackage{subfiles}
\usepackage{epstopdf}

\usepackage[normalem]{ulem}

\usepackage{tikz}
\usetikzlibrary{positioning}

\usepackage{blkarray}
\usepackage{multirow}

\newcommand{\be}{\begin{equation}}
\newcommand{\ee}{\end{equation}}
\newcommand{\bea}{\begin{eqnarray}}
\newcommand{\eea}{\end{eqnarray}}

\newcommand{\KL}{\langle\!\langle}
\newcommand{\KR}{\rangle\!\rangle}

\usepackage{tabularx}
\newcolumntype{L}{>{\raggedright\arraybackslash}X}
\usepackage{makecell}
\begin{document}
\preprint{APS/123-QED}

\title{Coulomb drag effect induced by the third cumulant of current}

\author{Artem Borin}
\affiliation{D\'epartement de Physique Th\'eorique, Universit\'e de Gen\`eve, CH-1211 Gen\`eve 4, Switzerland}
\author{Ines Safi}
\affiliation{Laboratoire de Physique des Solides (UMR 5802), CNRS-University Paris-Sud/Paris-Saclay, B\^at. 510, 91405 Orsay, France}
\author{Eugene Sukhorukov}
\affiliation{D\'epartement de Physique Th\'eorique, Universit\'e de Gen\`eve, CH-1211 Gen\`eve 4, Switzerland}

%
%
%


\date{\today}

\begin{abstract}
The Coulomb drag effect arises due to electron-electron interactions when two metallic conductors are placed in close vicinity to each other. It manifests itself as a charge current or  voltage drop induced in one of the conductors, if the current flows through the second one. Often it can be interpreted as an effect of rectification of the nonequilibrium {\em quantum} noise of current. Here, we investigate the Coulomb drag effect in mesoscopic electrical circuits and show that it can be mediated by {\em classical} fluctuations of the circuit collective mode. Moreover, by considering this phenomenon in the context of the full counting statistics of charge transport we demonstrate that not only the noise power, but also the third cumulant of current may contribute to the drag current. We discuss the situations, where this contribution becomes dominant.
\end{abstract}


\maketitle


\section{Introduction}
The Coulomb drag effect is the phenomenon observed in a system of two interacting conducting circuits, which manifests itself as a charge current or a voltage drop induced in a {\em drag} circuit, when a charge current flows through a {\em drive} circuit. \footnote{For a review, see B. N.  Narozhny and A. Levchenko, Rev. Mod. Phys. {\bf 88}, 025003 (2016), and references therein} It originates from the broken electron-hole symmetry and electron interactions, and therefore it is often studied in mesoscopic systems of reduced dimensionality, such as quantum wires\citep{exp1d1,exp1d2,exp1d3,exp1d4,exp1d5,the1d1,the1d2}, quantum dots\citep{expQD1,theQD1,theQD2,theQD3},  and   
quantum point contacts\citep{expqpc1,expqpc2,kamenev2}, where both effects are strongly pronounced. Its manifestation is particularly interesting in quantum conductors, where the electron-hole asymmetry is connected to the energy dependence of the transmission coefficients\cite{Blanter-Buttiker} and can be tuned by applying a gate voltage. In such systems the Coulomb drag can be viewed as an effect of rectification by the drag circuit of quantum noise of the drive circuit.\cite{kamenev2} Although known also in higher dimensions,\citep{kamenev1} this effect is more evident in low-dimensional systems,  where it can be used to measure the spectral density of the noise\citep{expQD1,theQD1} and its properties \citep{kamenev2}, as well as to probe fundamental fluctuation relations\citep{theQD2}. 

In all mentioned above examples the main contribution to the drag effect comes from the two-point correlation function $\langle\delta I(t)\delta I(0)\rangle$ of current fluctuations $\delta I$ in the quantum conductor at time scales of the order of the correlation time 
(typically given by one over the  voltage bias or temperature), where fluctuations are essentially quantum. From a broader perspective of the full counting statistics (FCS) of quantum conductors, \cite{Levitov} such a correlation function is a characteristic of the Gaussian noise.  To clarify this fact, let us consider the moment generating function
\begin{equation}
\label{moment}
Z(\lambda, t)=\sum_Q e^{i\lambda Q}P(Q, t), 
\end{equation}
 of the charge $Q$ transmitted through a quantum conductor during time $t$, and for simplicity take the Markovian (classical noise) limit, $t\gg \tau_c$,  where the short-time fluctuations contribute to the generator independently:  
\begin{equation}
\label{cumulant}
\ln[Z(\lambda, t)]={\cal H}(\lambda)t, \quad {\cal H}(\lambda)=\sum\limits_{n=1}^\infty\KL I^n\KR\frac{(i\lambda)^n}{n!}.
\end{equation}
Here $\KL I^n\KR$ are the current cumulants, the first three being the average current, $\KL I\KR=\langle I\rangle$, the zero-frequency noise power, $\KL I^2\KR=\int dt\langle\delta I(t)\delta I(0)\rangle $, and the third cumulant  $\KL I^3\KR=\int dt\int dt'\langle\delta I(t)\delta I(t')\delta I(0)\rangle$. Although the current cumulants enter the FCS generator ${\cal H}$ on equal footing, experimentally the high-order cumulants are much less accessible than the second one,  because in large systems their contributions to measured quantities (including the drag current) are suppressed due to the central limit theorem. The third current cumulant has been experimentally studied by explicitly collecting the statistics of the transferred charge\citep{3dmom1,3dmom2,3dmom3,3dmom4,3dmom5,3dmom6}, and by studying the weak asymmetry of the escape rate in Josephson junction threshold detectors\citep{3dmom7,3dmom8} with  respect to the current bias.

\begin{figure}[h]
	\centering
		\includegraphics[scale=0.9]{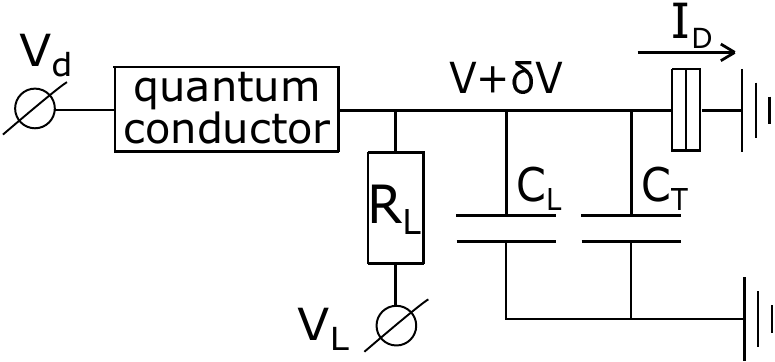} 
	\caption{The simplified electrical circuit for studying the Coulomb drag effect is shown. It consists of two parts: the drive circuit, containing a quantum conductor with the resistance $R_S$, which is the source of the current noise, and the drag circuit, containing a tunnel junction with the resistance $R_T$ and capacitance $C_T$, which serves as a detector of noise. The voltage bias $\Delta V= V_d-V$, applied to the quantum conductor, causes the average current $\langle I\rangle$ and nonequilibrium current fluctuations $\delta I$ through it. The circuit responds by the voltage fluctuations $\delta V$ accross the tunnel junction at the characteristic frequency $\omega_c=1/(RC)$ , where the circuit resistance is defined as $R^{-1}=R_L^{-1}+R_S^{-1}$, and the total capacitance is given by $C=C_L+C_T$. These fluctuations cause the drag current $I_D$ through the tunneling junction, which is calculated perturbatively in small $1/R_T$. The extra potential $V_L$ is applied to tune the circuit to the point $V=0$ in order to cancel the average bias acrross the junction. An example of an open circuit for the drag effect detection is discussed in Sec.\ \ref{secSet}.}
\label{fig:setup1}
\end{figure}

Alternatively, one can consider the Coulomb drag effect in mesoscopic circuits in the context of the noise detection physics,  where the drive circuit generates nonequilibrium noise, while the drag circuit plays the role of the detector. It turns out \cite{edwards} that the current through the tunnel junctions detector is expressed in tems of the correlation function, which in the long-time limit acquires the form $e^{i V t}Z(\lambda, t)$, where $ V$ is the voltage bias accross the junction, $Z(\lambda, t)$ is the moment generator (\ref{moment}), and the counting variable $\lambda$ plays the role of the effective coupling constant. Typically, the effective coupling between the drive and drag circuits is weak, $\lambda\ll 1$, which explains the suppression of cumulants of the order $n>2$. Note, however, that according to Eq.\ (\ref{cumulant}) the third cumulant of current, in contrast to the second one, simply shifts the voltage bias in this correlation function: $V\to V-\lambda^3\KL I^3\KR/6$. This leads to the idea, that Markovian (classical) odd cumulants of noise, being a nonequilibrium property of a quantum conductor (they vanish at zero bias) may propagate to the drag circuit and cause the DC drag current by shifting the bias. Being essentially classical, this phenomenon has to be differentiated from the drag effects studied so far, where the main contribution come from short time scales given by the correlation time of current fluctuations.\cite{kamenev2} However, this simple idea has a caveat that we explain below.

A simplified electrical circuit for detecting the drag effect is shown in Fig.\ \ref{fig:setup1}. (An alternative open circuit set-up is considered in Sec.\ \ref{secSet}.) It contains a quantum conductor that emits a nonequilibrium current noise, and a tunnel junction detector, where the drag current is induced.   The fluctuations of the current in the conductor, $\delta I$, do not propagate directly towards the detector: they are accumulated in the capacitor and lead to voltage fluctuations $\delta V$ across the tunnel junction. Current and voltage fluctuations are related by the solution of the Langevin equation:
\begin{equation}
\label{Ohm}
\delta V(\omega) =  Z(\omega) \delta I(\omega),\quad Z(\omega)=\frac{R}{1-i\omega/\omega_c},
\end{equation}
where $Z(\omega)$ is the impedance of the circuit, $\omega_c=1/(RC)$ is the circuit response frequency, the circuit resistance is defined as $R^{-1}=R_L^{-1}+R_S^{-1}$, and the total capacitance is given by $C=C_L+C_T$. At long times, $\omega_ct\gg 1$, i.e., at low frequencies, one  obtains $\delta V=R\delta I$, giving indeed $\KL V^3\KR=R^3\KL I^3\KR$, where $R$ plays the role of an effective coupling constant. However, for an Ohmic tunnel junction the main contribution comes from short time scales, $t\ll 1/\omega_c$, where, due to the prefactor $Z(\omega)$, the voltage fluctuations are suppressed 
as $1/\omega^2$ [see Eq.\ (\ref{Ohm})]. In Sec.\  \ref{WC} we rigorously show that this leads to complete cancellation of the drag effect from the classical noise in Ohmic tunnel junctions.\nocite{levch} \footnote{The quantum drag effect to third order in coupling has been studied in [\onlinecite{levch}]}  Therefore, we focus in the paper on the tunnel junctions with different nonlinear I-V characteristic, find the drag current $I_D$ perturbatively in small $1/R_T$, and express it in terms of the Markovian third cumulant  of current of the quantum conductor, $\KL I^3\KR$,  and the circuit parameters.

This paper is organized as follows. In Sec.\ \ref{secMod} we use the $P(E)$-theory of tunneling\citep{Nazarov1,*Naz} to derive the expression for the drag current in terms of the I-V characteristic of the tunnel junction.  Then, in Sec.\ \ref{WC} we apply the weak coupling expansion to formally express the drag current in terms of the cumulants of the current of the quantum conductor.  In Sec.\ \ref{secCalc} we separately consider the drag effect in tunnel junctions with analytical and nonanalytical I-V characteristic. In Sec.\ \ref{secSet} we investigate the drag effect in the open circuit set-up (see Fig.\ \ref{fig:setup}), which is experimentally more relevant. Finally, in Appendix\ \ref{secSPI} we use the stochastic path integral (SPI) technique\citep{stochpath,*stochpath2} to derive the second and third voltage cumulants in terms of the current cumulants in the quantum conductor.

\section{P(E)-theory of tunneling and the drag effect}
\label{secMod}
 
In this and next section we closely follow Refs.\ [\onlinecite{edwards}], [\onlinecite{Safi1}] and [\onlinecite{Safi2}]. We consider a tunnel junction in the presence of the noise of the collective mode, propagating in an electrical circuit from a quantum conductor, as shown in the Fig.\ \ref{fig:setup1}, and apply the $P(E)$ theory of tunneling \citep{Nazarov1,*Naz}  to evaluate the current in the tunnel junction induced by this noise. The advantage of this approach is that to the leading (second) order in tunneling there is no need to specify the Hamiltonian of the reservoirs of the tunnel junction to derive the expression for the drag current: an arbitrary disorder and interactions can be included. The electron tunneling  is described by the  Hamiltonian (throughout the paper we use unites, where $|e|=\hbar=1$)
\begin{equation}
\label{H_T}
H_T=A + A^\dagger ,
\end{equation} 
where the tunneling operator $A$ transfers an electron from the left to the right reservoir. According to the tunneling Hamiltonian approach \cite{Mahan}, the tunneling current operator  
can be defined as $I\equiv -dN_R/dt=i(A-A^\dagger) $, where $N_R$ is the number of electrons in the right reservoir. Thus, in the leading order in  tunnelling  the average value of the tunneling current $I_T\equiv \langle I\rangle$ is given by:
\begin{equation}
\label{tunnel_cur}
I_T(V)=\int dte^{iVt}\langle[A(t),A^{\dagger}(0)]\rangle,
\end{equation}
where $V$ is the applied voltage bias, and the average is evaluated with respect to the equilibrium state: $\langle\ldots\rangle=\sum_n\rho_n\langle n|\ldots|n\rangle$, and $\rho_n\propto e^{-E_n/T}$ with $T$ being the bath temperature.
In the absence of noise in the circuit, this expression give the bare I-V characteristic $I_0(V)$ of the junction, which, according to the structure of Eq.\ (\ref{tunnel_cur}), can also be presented
as 
\begin{equation}
\label{LR-RL}
I_0(V)=I_{LR}(V)-I_{RL}(V),
\end{equation}
where the two terms on the right hand side differ by the direction of electron tunneling.

In the next step, we account for coupling of the junction to noise by  substituting \citep{Nazarov1,*Naz}  
\begin{equation}
A\to e^{i\phi}A,\quad A^\dagger\to e^{-i\phi}A^\dagger,
\label{subst}
\end{equation} 
where  the operator $e^{i\phi}$ increases the charge on the capacitor by $1$, which can be expressed as $[\phi,Q]=i$. Then, the charge Hamiltonian $H_C=Q^2/2C$ generates the equation of motion:
\begin{equation}
\label{eqmo}
\dot\phi=Q/C=\delta V,
\end{equation}
where $\delta V$ is fluctuating part of the voltage across the tunnel junction.
After substituting the operator $A$ from Eq.\ (\ref{subst}) into Eq.\ (\ref{tunnel_cur}) and using Eq.\ (\ref{LR-RL}), we arrive at the following expression for the tunneling current: 
\begin{align}
I_T(V)=\int_{-\infty}^{\infty}d\omega [P_{LR}(\omega)I_{LR}(V-\omega)\hspace*{1cm}\nonumber\\
  -P_{RL}(-\omega)I_{RL}(V-\omega)],
\label{drag_current}
\end{align}
where 
\begin{eqnarray}
\label{probbos1}
P_{LR}(\omega)=\frac{1}{2\pi}\int dt e^{i\omega t} \langle e^{i\phi(t)}e^{-i\phi(0)}\rangle , \\
\label{probbos2}
P_{RL}(\omega)=\frac{1}{2\pi}\int dt e^{i\omega t} \langle e^{-i\phi(t)}e^{i\phi(0)}\rangle  
\end{eqnarray}
are  the probabilities of absorbing the energy $\omega$ from  the circuit which depend on the direction of tunneling.

We are now in the position to discuss how the drag current vanishes at equilibrium. Therefore, we set $V=0$ in Eq.\ (\ref{drag_current}) and assume, that the circuit is in equilibrium at the temperature $T_C$. We then apply the spectral decomposition to the probabilities (\ref{probbos1}) and (\ref{probbos2}), and write $P_{LR}(\omega)=\sum_{nm}\rho_{C,n}|\langle n|e^{i\phi}| m\rangle|^2\delta(\omega+E_n-E_m)$. By comparing this expression to the similar result for $P_{RL}(\omega)$ and using the equilibrium weights $\rho_{C,n}\propto e^{-E_n/T_C}$, we arrive at the detailed balance equation: 
\begin{equation}
P_{RL}(\omega)=P_{LR}(-\omega)e^{\omega/T_C}.
\end{equation} 
Assuming now that the detector tunnel junction is at equilibrium with the temperature $T_D$ and applying the spectral decomposition to the equations (\ref{tunnel_cur}) and (\ref{LR-RL}), we obtain 
\begin{equation}
I_{RL}(\omega)=e^{-\omega/T_D}I_{LR}(\omega).
\end{equation}  
Using these two detailed balance equations in the equation (\ref{drag_current}), we arrive at the following result:
\begin{equation}
I_T(0)=\int_{-\infty}^{\infty}d\omega P_{LR}(\omega)I_{LR}(-\omega)\left[1-e^{-\omega/T_C}e^{\omega/T_D}\right].
\label{equi}
\end{equation}

\nocite{edwards}

We note that all the available frequencies, including quantum fluctuations, contribute to the total current (\ref{equi}), and all these contributions cancel at the global equilibrium $T_C=T_D$. Knowing this fact, in what follows we assume that the main contribution to the drag current comes from nonequilibrium fluctuations at low frequencies, which can be considered classical (Markovian). In other words, we assume that the phase operator $\phi(t)$ commutes with itself at different times, so that the expressions (\ref{probbos1}) and (\ref{probbos2})
simplify, and introducing the new notation $P(\omega)=P_{LR}(\omega)$ one can write
\begin{equation}
\label{Pomega}
P(\omega)=\frac{1}{2\pi}\int dt e^{i\omega t} \langle e^{i[\phi(t)-\phi(0)]}\rangle,
\end{equation}
while $P_{RL}(\omega)=P(-\omega)$. Applying these simplifications in Eq.\ (\ref{drag_current}), using Eq.\ (\ref{LR-RL}), and introducing yet another notation $I_D$ for the drag current, we arrive at its final general form for the case of classical noise:
\begin{equation}
I_D\equiv I_T(0),\quad I_T(V)=\int_{-\infty}^{\infty}d\omega I_0(\omega)P(V-\omega),
\label{final-drag}
\end{equation}  
where, we recall, $I_0(V)$ is the bare tunneling current, not affected by fluctuations in the circuit.

\section{Weak coupling expansion}
\label{WC}

We further assume that coupling of the junction to the system, described by Eqs.\  (\ref{Ohm}), is weak, i.e., $R\ll 1$,  and expand the probability $P(\omega)$ in cumulants of the phase  $\phi$ to the third order:
\begin{eqnarray}
\label{prob}
P(\omega)=\frac{1}{2\pi}\int dt e^{i\omega t-J_2(t)-iJ_3(t)}, 
\end{eqnarray}
where the cumulants are given by
\begin{eqnarray}
\label{cumulants_def}
J_2(t)=\frac{1}{2}\langle [\phi(t)-\phi(0)]^2\rangle, \\
\label{cumulants_def2}
J_3(t)=\frac{1}{6}\langle [\phi(t)-\phi(0)]^3\rangle .
\end{eqnarray}
Here we assumed the semi-classical (Markovian) noise limit, $\Delta V/\omega_c\gg 1$ with $\Delta V=V_d-V$ being the bias over the quantum conductor, and neglect quantum corrections.\footnote{Note that $J_3$ is smaller than $J_2$ by the dimensionless coupling constant $R\ll 1$ (in unites where $|e|=\hbar=1$), therefore one should consider the quantum correction to $J_2$. However, it has been shown in Ref.\ [\onlinecite{edwards}], that  according to the Kubo linear response formula it is simply given by the differential conductance of the noise source, which is an equilibrium property. Therefore, it does not contribute to the drag effect.}  The correlators $J_2$ and $J_3$ are evaluated in Appendix A. In the next section we will consider the cases of a slow and  fast circuit, where these correlators have to be taken in the short-time,  $\omega_c|t|\ll 1$, and long-time, $\omega_c|t|\gg 1$, limits, respectively.
 
In the short-time limit, $\omega_c|t|\ll 1$, one finds 
\begin{equation}
\label{short-time}
J_2(t) = \frac{K^{(s)}_2t^2}{2},\quad
J_3(t)= \frac{K^{(s)}_3t^3}{6},
\end{equation}
where the coefficients represent the second and third cumulant of ``instant'' (at equal times) fluctuations of the potential: $K^{(s)}_m=\langle(\delta V)^m\rangle$, $m=2,3$. Using Eqs.\ (\ref{Ohm}), they can be expressed in terms of the second and the third cumulant of the current in the quantum conductor at zero frequencies:
\begin{equation}
\label{short-time2}
K_2^{(s)} = (R/2C)\KL I^2\KR, \quad
K_3^{(s)}=(R/3C^2)\KL I^3\KR_{\rm tot},
\end{equation}
where the total third cumulant reads
\begin{equation}
\label{cascade_s}
\KL I^3\KR_{\rm tot}=\KL I^3\KR-\frac{3\KL I^2\KR^2\partial_Q\omega_c}{2\omega_c^2}
+\frac{3\KL I^2\KR\partial_Q\KL I^2\KR}{2\omega_c} .
\end{equation}
Note that the second and third terms  represent circuit cascade corrections due to the nonlinear and environmental effects, respectively. These corrections are specific to the regime of a slow circuit, and they can be found  using the SPI method, as demonstrated in Appendix A (see also Ref.\ [\onlinecite{edwards}]).

In the long-time limit, $\omega_c|t|\gg 1$, one obtains
\begin{equation}
\label{long-time}
J_2(t) =  \frac{K^{(f)}_2|t|}{2},\quad
J_3(t)=\frac{K^{(f)}_3t}{6}, 
\end{equation}
where the coefficients $K_m^{(f)}$, $m=2,3$ can be read off the Eqs.\ (\ref{Ohm})  by replacing $Z(\omega)\to R$,
\begin{equation}
K^{(f)}_2 = R^2\KL I^2\KR,\quad
K^{(f)}_3 =R^3\KL I^3\KR_{\rm tot}.
\label{cf}
\end{equation}
However, as in the case of the slow circuit, the cascade corrections for a fast circuit can be obtained by the SPI method:
\begin{equation}
\label{cascade_f}
\KL I^3\KR_{\rm tot}=\KL I^3\KR-\frac{6\KL I^2\KR^2\partial_Q\omega_c}{\omega_c^2}
+\frac{3\KL I^2\KR\partial_Q\KL I^2\KR}{\omega_c},
\end{equation}
where, again, the second and third term  represent circuit cascade corrections due to the nonlinear and environmental effects, respectively.
The environmental effects in the third cumulant have been experimentally studied in Refs.\ [\onlinecite{3dmom1}] and [\onlinecite{3dmom4}].

By using  Eqs.\ (\ref{final-drag}) and (\ref{prob}), the drag current can be written as
\begin{equation}
I_D
=\int \frac{dtd\omega}{2\pi}I_0(\omega)e^{i\omega t - J_2(t)+iJ_3(t)}\,.
\label{IDder}
\end{equation} 
The way the third current cumulant enters this expression suggests that it may have a similar effect on the tunnel junction  as the DC voltage bias, i.e., it may cause a drag current. Indeed, in the long-times limit \eqref{long-time} the third current cumulant enters as a linear in time phase shift, i.e., it adds to the voltage bias, so one expects a finite current even at zero voltage.  However, it turns out,  that for an Ohmic tunnel junction, $I_0(\omega) \propto \omega $, and for a classical noise considered here the drag current vanishes. Indeed, in this case the integral over $\omega$ in Eq.\ (\ref{IDder}) imposes the $t\to 0$ limit, and one obtains:
\begin{equation}
I_D\propto \partial_t [-J_2(t)+iJ_3(t)]_{t=0}=0,
\end{equation}
according to  Eq.\ (\ref{short-time}).
Therefore, in the rest of the paper, we consider the drag current in tunnel junctions with different nonlinear I-V characteristic.

\section{Drag current for tunnel junctions with nonlinear I-V characteristic}
\label{secCalc}

In this section we evaluate the drag current for two types of nonlinearities in the tunnel junction. Namely, in Sec.~\ref{anal} we consider the analytical regime, $I_0(V)= \sum_n g_nV^n$, where $n=1,2,\ldots$, while in Sec.~\ref{non-anal} we investigate the nonanalytical regime, $I_0(V)= g_\alpha V|V|^{\alpha-1}$ for noninteger $\alpha$. However, before proceeding with calculations, one needs to check that the main contribution to the integral in Eq.\ (\ref{IDder}) comes from frequencies smaller than $\Delta V$  to ensure that our classical noise approximation still applies (i.e., the noise source $\delta I$ can be considered Markovian). Since $P(\omega)$, given by Eqs.\  (\ref{prob})-(\ref{cumulants_def2}), is already taken in the classical limit, it is sufficient to check that the integral in Eq.\ (\ref{final-drag}) does not diverge at infinity for $V=0$. For doing so, let us consider the asymptotic behavior of Eq.\ (\ref{prob}) for frequencies $\omega\gg \max\{\omega_c,(K_2^{(s)})^{1/2}\}$. Using the short-time dependence of the cumulant \eqref{J2} we find that the even (odd) part of $P(\omega)$ scales as $\omega^{-\gamma}$ with $\gamma = 4$ ($\gamma = 5$). Therefore, as long as $n\leq 3$ in the analytical regime and $\alpha\leq 4$ in the nonanalytical regime the classical noise approximation is valid. Outside this parameter range, either the quantum character of the noise or  high-frequency cutoff of $I_0(V)$ should be taken into account, which is beyond the scope of this paper.

\subsection{Analytical regime}
\label{anal}

Assuming the analytical I-V dependence, 
\begin{equation}
\label{analit}
I_0(V)= \sum_n g_nV^n,\quad n=1,2,\ldots,
\end{equation}
 and that the integral (\ref{IDder}) converges, we first perform the integration over $\omega$, and then remove resulting delta-functions by the integral over time,
\begin{equation}
\label{intermed}
I_D=\sum_{n=1}^{\infty}i^ng_n\partial^n_{t}e^{-J_2+iJ_3}|_{t\to 0}=g_2K_2^{(s)}+g_3K_3^{(s)},
\end{equation}
where we kept only the first two terms of the expansion, since higher-order terms are small due to the weak coupling regime. Alternatively, one can derive this expression by approximating $\phi(t)-\phi(0)\approx\delta V t$ at short times in Eq.\ (\ref{Pomega}) and using Eq.\ (\ref{final-drag}). This approximation holds  up to the third order in $\delta V$ and gives 
\begin{equation}
\label{rect}
I_D=\langle I_0(\delta V)\rangle.
\end{equation}
Substituting here $I_0$ from Eq.\ (\ref{analit}), one arrives at Eq.\  (\ref{intermed}). Thus, the drag current in this regime is due to the {\em rectification} of the instant fluctuations of the potential. Using Eq.\ (\ref{short-time2}), we arrive at the result:
\begin{equation}
\label{res1}
I_D=\frac{g_2}{2}R^2\omega_c\KL I^2\KR+\frac{g_3}{3}R^3\omega_c^2\KL I^3\KR_{\rm tot},
\end{equation}
The third cumulant contribution  is an even function of the source current and  can be measured by changing its direction, even though it is the subdominant contribution to the drag current.

In what follows, we  discuss our finding in the context of earlier published results. This concerns the noise rectification effect. We note that the first term in Eq.\ (\ref{res1}) has a simple structure: it is proportional to the product of the noise power $\KL I^2\KR$ and the circuit response frequency $\omega_c$. Estimating the source noise as $\KL I^2\KR\propto \Delta V$, where the $\Delta V$ is the voltage bias applied to the quantum conductor, we conclude that the noise rectification contribution  scales as  $\omega_c\Delta V$.
The same structure can be found in the drag current derived in Ref.\   [\onlinecite{kamenev2}] in the quantum regime, $\omega_c\gg\Omega=\max(\Delta V, T)$, where $T$ is the temperature of the system. In this case the frequency integrals are limited by $\Omega$. In the shot noise limit, $\Delta V\gg T$  the noise power is proportional to the bias applied to the quantum conductor, $ \Delta V$, while the cutoff frequency is determined by the same  scale (as discussed above), which leads to the following result  $I_D\propto \Delta V^2$.

Close to equilibrium, $\Delta V\ll T$, the drag effect originates from an even component of the nonlinearity in the I-V characteristic of the quantum conductor that scales as $\Delta V^2$. \cite{kamenev2} In this case the noise power scales as  $\KL I^2\KR\propto \Delta V T$, while the frequency cutoff is given by the temperature, resulting in $I_D \propto \Delta V T^2$, which becomes $I_D \propto \Delta V T\omega_c$ in the case of classical noise. Interestingly, this contribution to the drag current depends on the direction of the source current, and thus it may compete with the third cumulant contribution. Therefore, we propose to do measurements in  the regime, where the I-V characteristic of the mesoscopic conductor is an odd function.

\subsection{Nonanalytical regime}
\label{non-anal}

In this section we consider tunnel junctions with nonanalytical I-V characteristic of the form
\begin{equation}
\label{nonanalit}
I_0(V)=g_{\alpha} V|V|^{\alpha-1},
\end{equation}
  where $\alpha$ is noninteger number, and $g_\alpha$ is an arbitrary constant. Such I-V characteristic is typical for systems with interactions, e.g., Luttinger liquids or disordered systems. Since $I(V)$ is an odd function of the voltage bias $V$, only the third current cumulant contributes to the drag effect, as one can easily see from Eq.\ (\ref{IDder}). Due to  weak coupling, and since the time integral in Eq.\ (\ref{prob}) is limited by $J_2$, one can expand the exponential function in the integral in small $J_3$: $P(\omega)=P_0(\omega)+\delta P(\omega)$, where
\begin{equation}
\label{prob2}
\delta P(\omega) = (i/2\pi)\int dt e^{i\omega t - J_2(t)}J_3(t),
\end{equation}
In contrast to the analytical regime, here one should separately consider the cases of the slow circuit, $\omega_c\ll R^2\KL I^2\KR$, and of the fast circuit, $\omega_c\gg R^2\KL I^2\KR$. Note that in the latter case $\omega_c$ is still bound from above, because the circuit response should be slower then the correlation time of the noise: $\omega_c \ll \Delta V$. This is consistent with the requirement of weak coupling $R\ll 1$, since $\KL I^2\KR\sim \Delta V$.

\subsubsection{Slow circuit, $\omega_c\ll R^2\KL I^2\KR$.}

In this case the contribution to the integral in Eq.\ \eqref{prob2} comes from times $t\ll 1/ \omega_c$, therefore we use the short-time limit  \eqref{short-time} for the phase correlation functions.  Substituting these expressions into Eq.\ \eqref{prob2}, we obtain 
\begin{equation}
\label{prob_slow}
\delta P(\omega)=-\frac{e^{-\frac{\omega^2}{2K^{(s)}_2}}}{\sqrt{2\pi K^{(s)}_2}} 
\left\{\frac{\omega K^{(s)}_3}{2(K^{(s)}_2)^2}-\frac{\omega^3 K^{(s)}_3}{6 (K^{(s)}_2)^3} \right\}.
\end{equation}
Substituting this expression for the correction to the  probability distribution function  along with nonanalytical I-V characteristic into the  Eq.\ \eqref{IDder},  we arrive at the result for the drag current for $-2<\alpha<4$
\begin{equation}
\label{slow_nonanal}
I_D=\frac{ 2^{(\alpha-1)/2}(\alpha-1) g_\alpha}{3\sqrt{\pi}}\Gamma\left(\frac{2+\alpha}{2}\right)
[K^{(s)}_2]^{(\alpha-3)/2}K^{(s)}_3,
\end{equation}
where the correlation functions $K^{(s)}_m$, $m=2,3$, are expressed in terms of the current cumulants in Eqs.\ (\ref{short-time2}) and (\ref{cascade_s}). This expression correctly reproduces the above results for the Ohmic ($\alpha=1$) and cubic ($\alpha=3$) terms in I-V characteristic of the tunnel junction [see Eq.\ (\ref{res1})].

For $\alpha<-2$ the integral in Eq.\ \eqref{IDder} becomes divergent at small frequencies. Introducing the infrared cutoff, $\omega_0$, we express the drag current as:  
\begin{equation}
I_D\propto g_\alpha K_3^{(s)}/[(K_2^{(s)})^{5/2}\omega_0^{-\alpha-2}].
\end{equation} 
 Interestingly, for the case of a nonanalytical  I-V characteristic of a tunnel junction, the drag current  depends both on the second and the third cumulants, in contrast to the case of the analytical nonlinearity. It is clear that this result can not be obtained perturbatively in noise power, since $K_2^{(s)}$ enters this expression nonanalytically.

\subsubsection{Fast circuit, $\omega_c\gg R^2\KL I^2\KR$.}

Taking into account the result (\ref{long-time}) and using Eq.\ (\ref{prob2}) we arrive at the following expression for the third cumulant correction to the probability distribution function:
\begin{equation}
\delta P(\omega)=\frac{i}{2\pi}\int dt e^{i\omega t-K_2^{(f)}|t|}J_3(t),
\end{equation}
where the correlator $K_2^{(f)}$ is given by Eq.\ (\ref{cf}). (This result holds up to small relative  corrections of the order of $R^2\KL I^2\KR/\omega_c$). We first concentrate on the case $1<\alpha<4$, where an interesting situation arises: the drag current is determined by neither the  short-time nor the long-time limit of $J_3(t)$. Consequently, it acquires unusual nonlinear and environmental cascade corrections that have not been discussed in literature.  Straightforward calculations lead to the following result:
\begin{align}
I_D=\frac{g_\alpha R^3}{\pi}\bigg\{C(\alpha) \left(\omega_c^{\alpha-1} \KL I^3\KR -\frac{3}{2}\omega_c^{\alpha-3} \KL I^2\KR^2\partial_Q\omega_c  \right)
\nonumber \\  +2F(\alpha)\omega_c^{\alpha-2} \KL I^2\KR\partial_Q \KL I^2\KR\bigg\},
\label{fast_nonanal}
\end{align}
where $C(\alpha)=\int^{\infty}_0dx x^{\alpha-1}/(x^2+1)(x^2+4)$ and $F(\alpha)=\int^{\infty}_0dx x^{\alpha-1}/(x^2+1)^2$. This expression is obtained up to corrections of the order of $[R^2\KL I^2\KR/\omega_c]^{\alpha-1}$. Note that for $\alpha=3$ it agrees with the third cumulant contribution in Eq.\ (\ref{res1}). 

In contrast, for $-2<\alpha<1$, the drag current becomes determined solely by the long-time behaviour of $J_3(t)$, therefore it can be expressed in terms of the correlators (\ref{cf}) and (\ref{cascade_f}),
\begin{equation}
\label{fast_nonanal1}
I_D=-\frac{4g_\alpha}{\pi}F(\alpha+2) [K^{(f)}_2]^{\alpha-1} K^{(f)}_3, 
\end{equation}
where the function $F$ is introduced above. For 
 $\alpha<-2$, we regularise I-V characteristic at small voltages by the cutoff $\omega_0$. Then the drag current in this case has the form given by  Eq.\ \eqref{fast_nonanal1} after the replacement $F(\alpha+2)\to (\omega_0/\omega_c)^{\alpha+2}$. 
 
 Finally, we note that the drag current obtained in this subsection originates from the long-time behaviour of $J_3(t)$, which results in the shift of the voltage bias, discussed in the introduction. Therefore, it has to be differentiated with the noise rectification effect arising at short time scales and discussed in Sec.\ \ref{anal}.

\section{Drag effect in the open circuit setup}
\label{secSet}

\begin{figure}[h]
	\centering
		\includegraphics[scale=0.8]{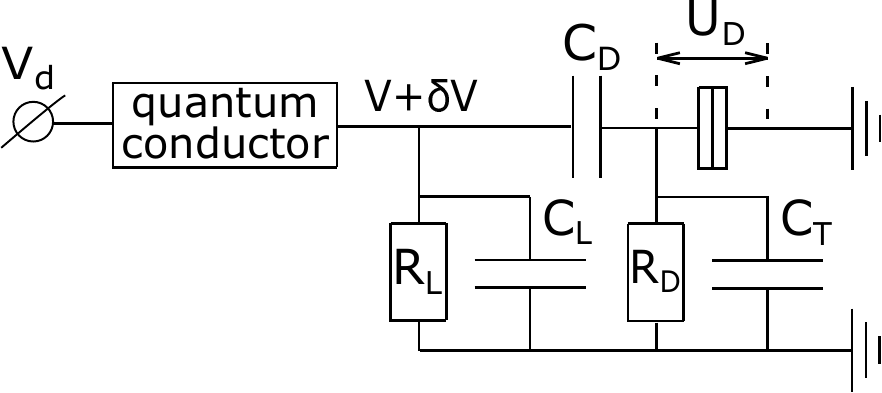} 
	\caption{An example of an open circuit for studying the Coulomb drag effect is shown. Compared to the circuit shown in Fig.\ \ref{fig:setup1}, an extra capacitor $C_D$ is added in order to filter out the DC component of the voltage, $V$, as well as the low-frequency part of fluctuations $\delta V$. The high-frequency part of  fluctuations propagates towards the detector part of the circuit and causes the drag voltage $U_D$ across the tunnel junction. One can use the additional shunt resistor $R_D$ to controllably access the nonlinear regime of the tunnel junction. The relation between the drag voltage $U_D$ in the open circuit and the drag current $I_D$ in the circuit shown in Fig.\ \ref{fig:setup1} is studied in Sec.\ \ref{secSet}.}
\label{fig:setup}
\end{figure}

In  Sec.\ \ref{secCalc} we derived the  drag current $I_D$ for the most elementary setup shown in  Fig.\ \ref{fig:setup1}, where the tunnel junction is electrically connected to the circuit. The disadvantage of such connection is that it might be difficult to measure the drag effect due to the third cumulant of the current in the background of the nonzero average DC current contribution. Fortunately, our results can be easily modified for the case of an experimentally more relevant setup shown in Fig.\ \ref{fig:setup}, where the drag voltage $U_D$ is measured, and where the DC component of the average current is filtered out.

First, we note that the role of the capacitor $C_D$ in the setup in Fig.\ \ref{fig:setup} is to filter out the DC component of the bias, $V$. However, one has to be sure that the largest part of fluctuations $\delta V$ still propagates towards the tunnel junction. This is the case when, on one hand, the detector circuit does not screen the fluctuations and, on the other hand, only a small part of the voltage drops across the capacitor $C_D$ at relevant frequencies. The former holds if the impedance of the detector circuit
\begin{equation}
Z_D(\omega)=\frac{i}{\omega C_D}\frac{1-i\omega(C_D+C_T)\tilde R}{1-i\omega C_T \tilde R},
\end{equation}
where $\tilde{R}^{-1}=R_T^{-1}+R_D^{-1}$, is large compared to the impedance  (\ref{Ohm}) of the drive circuit $Z(\omega)$ at the characteristic frequencies of fluctuations $\omega_c = (RC)^{-1}$. The letter condition holds if the impedance of the detector capacitor $i/(\omega C_D)$ is small compared  to the impedance of the rest of the detector circuit $(1/\tilde R-i\omega C_T)^{-1}$ at frequencies of the order of $\omega_c$. The two conditions are satisfied simultaneously, if (i) $\tilde RC_T\ll RC$, $\tilde RC_D\gg RC$, and $\tilde R\gg R$, or, alternatively, (ii) $\tilde RC_T\ll RC$, $C_D\gg C_T$, and $C\gg C_T$. These conditions imply that the detector is noninvasive and that our previous results for the drag current hold.

The drag voltage is determined by the condition
\begin{equation}
\label{UD}
I_T(U_D) +\frac{U_D}{R_D}=0,
\end{equation}
where $I_T$ is given by  Eq.\ (\ref{final-drag}). This equation 
follows from  Kirchhoff's law and the fact, that the DC current through the open circuit vanishes.   In the case of the tunnel junction with an analytical I-V characteristic [see Eq.\ (\ref{analit})],  Eq.\ \eqref{UD} is solved trivially, giving
\begin{equation}
\label{UD_anal}
U_D=  -\frac{R_DR_T}{R_D+R_T}  I_D.
\end{equation}
Note that the tunneling conductance $1/R_T$ arising here is nothing but the expansion coefficient $g_1$ in Eq.\ (\ref{analit}). 

In the case of a nonanalytical I-V characteristic (\ref{nonanalit}), one should distinguish between two limits depending on how the value of the drag voltage $U_D$ compares to the width $\Gamma$ of the distribution $P(\omega)$, which can be estimated as
\begin{equation}
\label{Gamma}
\Gamma = R^2 \KL I^2\KR\min \left(1,\frac{1}{R}\sqrt{\frac{\omega_c}{\KL I^2\KR}}\right).
\end{equation} 
If the width $\Gamma$ of the probability distribution is small compared to the drag voltage $U_D$, then Eq.\ (\ref{UD}) can be solved by expanding $I_0(\omega)$ in (\ref{final-drag}) around $\omega=U_D$, givin the relation
\begin{equation}
\label{xxx}
I_0(U_D)+U_D/R_D=-I_D,
\end{equation}
where $I_0$ is given by  Eq.\ (\ref{nonanalit}). 

If the first or second term on the left hand side of the equation (\ref{xxx}) dominates, one obtains $U_D=-|I_D/g_\alpha|^{1/\alpha}$ or $U_D=-R_DI_D$, respectively. Note that in this regime the expression for the drag current (\ref{res1}) still applies. However, the coefficients are expressed in terms of $U_D$, namely, $g_2=(g_\alpha/2)\alpha(\alpha-1)U_D^{\alpha-2}$ and $g_3=(g_\alpha/6)\alpha(\alpha-1)(\alpha-2)U_D^{\alpha-3}$. For instance, if $1/R_D=0$ and the second cumulant contribution to the drag current dominates, one has $|U_D|\propto R[\omega_c\KL I^2\KR]^{1/2}$, while for the case, where the third cumulant dominates, one gets $|U_D|\propto R[\omega_c^2\KL I^3\KR]^{1/3}$. When comparing $U_D$ to $\Gamma$ in Eq.\ (\ref{Gamma}), we see that for the fast circuit, $\omega_c\gg R^2\KL I^2\KR$,  the regime $U_D\gg\Gamma$ is indeed  realized.
However, for the slow circuit, $\omega_c\ll R^2\KL I^2\KR$, $U_D$ is of the order of $\Gamma$ or smaller. For the finite shunt conductance $1/R_D$ the second term in Eq.\ (\ref{xxx}) may start dominating. However, this may be compensated by even smaller values of $U_D$.

This brings us to the limit $\Gamma\gg U_D$. Expanding $I_T$ in Eq.\ \eqref{final-drag} with respect to the small $V=U_D$ and using Eq.\ (\ref{UD}), one again arrives at Eq.\ (\ref{UD_anal}). However, now the tunneling conductance is given by the following expression
\begin{equation}
R^{-1}_T = \int \partial_\omega I_0(\omega)P(-\omega) d\omega,
\end{equation}
where we integrated by parts. Thus, the noise simply smears out the singular I-V characteristic (\ref{nonanalit}) at voltages of the order of $\Gamma$, and one can estimate $1/R_T\sim g_\alpha\Gamma^{\alpha-1}$.
Since in this case the tunneling resistance depends on the properties of both the tunneling junction and the noise, it is convenient  to shunt the tunnel junction by $R_D\ll R_T$, so that
\begin{equation}
\label{xxxx}
U_D = - R_DI_D ,
\end{equation}
further lowering the drag voltage to values $U_D\ll\Gamma$. In this regime the results of  Sec.\ \ref{non-anal} for the drag current apply. 

\section{Summary}

It is natural to think of the Coulomb drag effect as resulting from the friction between electron systems of two adjacent conductors due to electron-electron scattering. It can be caused either by the direct Coulomb interaction, or by the exchange of virtual excitations, such as plasmons or phonons. However, in the case of the Coulomb drag in mesoscopic electrical  circuits it is more appropriate to think of the noise rectification effect, since the drag is mediated by the collective mode, such as a potential on a capacitor. Nevertheless, in the quantum regime, where the characteristic circuit response frequency $\omega_c$ is much larger than the effective noise temperature $\Omega=\max(\Delta V, T)$,\cite{kamenev2} one can still think that electrons of the drive circuit ``push'' electrons in the drag circuit thereby creating the drag current or voltage, because the circuit reacts to current fluctuations in a quantum conductor almost immediately. In this paper we consider the opposite regime, $\omega_c\ll \Omega$, and study the Coulomb drag effect in mesoscopic circuits mediated by the classical noise of a collective mode. This allows us to put our analysis in the context of the FCS\cite{Levitov} and to investigate the drag effect due to the Markovian (frequency independent) third cumulant of the current. The interest to the third current cumulant is motivated by the fact that this is essentially a nonequilibrium and non-Gaussian component of current noise. 

We consider a simple mesoscopic circuit, shown in Fig.\ \ref{fig:setup1} (and its experimentally more relevant modification in Fig.\ \ref{fig:setup}). It contains a quantum conductor, the source of noise, and a tunnel junction detector, where the drag current is induced. We evaluate the drag current perturbatively in the tunneling Hamiltonian using the $P(E)$-theory of tunneling\citep{Nazarov1,*Naz} and express it in terms of the second and third current cumulants, assuming weak coupling of the detector to the circuit. For doing so, we apply the SPI technique,\citep{stochpath,*stochpath2} the functional method of solving the circuit Langevin equations, which allows one to find circuit cascade corrections to high-order current cumulants. We find that, surprisingly, the drag current vanishes in the case of an Ohmic tunnel junction detector. Therefore, we concentrate on the drag effect induced in a tunnel junction detector with a nonlinear I-V characteristic.  

It is important to distinguish nonlinear I-V characteristic of the two sorts: the relatively smooth analytical I-V curve, that can be expanded in the voltage bias $V$ around $V=0$,  and I-V curve essentially nonanalytical at $V=0$ point, as in the case of various kind of zero-bias anomaly effects. Thanks to the tunneling Hamiltonian approach used in the paper, there is no need to specify the reason for such nonanalyticity. In the former case the contribution to the drag effect comes from short (but still Markovian) times scales, and the result takes a simple form (\ref{res1}). In the weak coupling regime considered in the paper the second cumulant contribution to the drag current dominates. However, it is an even function of the source current, therefore, the third cumulant contribution can be singled out by changing the direction of the current. The case of a nonanalytical I-V characteristic is special in the sense that not only the drag current is different for slow (\ref{slow_nonanal}) and fast (\ref{fast_nonanal1}) circuit, but also there is a regime, where the drag current (\ref{fast_nonanal}) acquires  contributions from different time scales. In this case it contains cascade corrections that have not been discussed in literature.   

Finally, we consider the drag effect in an open circuit (see Fig.\ \ref{fig:setup}), which is experimentally more relevant, because in this case there is no need to extract the drag current from the background contribution due to the DC voltage bias. Instead, one can measure the drag voltage induced across the tunnel junction and use the results (\ref{UD_anal}), 
(\ref{xxx}), and (\ref{xxxx}) to express it in terms of the ``bare'' drag current found in the paper.

\begin{acknowledgments}
We thank M.\ Reznikov for fruitful discussions. This work was supported by the Swiss National Science Foundation

\end{acknowledgments}

\appendix

\section{Stochastic path integral}
\label{secSPI}
Although, we need to find the correlation functions (\ref{cumulants_def}) and (\ref{cumulants_def2}) of the field $\phi$, it turns out to be  convenient to start directly with the generating function 
\begin{equation}
\label{gen_func}
Z(\chi) = \langle e^{\chi(\phi(t)-\phi(0))}\rangle,
\end{equation}
and evaluate it up to the third order in $\chi$ in the exponent using the functional method.
Since we are interested in the low frequency limit, where the field $\phi$ can be considered classical, we apply the SPI technique, \citep{stochpath,*stochpath2} which correctly implements averaging in (\ref{gen_func}) over solutions of the Langevin equation (\ref{Ohm}). Given the relation (\ref{eqmo}) of the field $\phi$ to the charge on the capacitor $Q$, we write Eq.\ \eqref{gen_func} as
\begin{eqnarray}
\label{SPI}
Z(\chi)&=&\int\!\! \mathcal{D}Q\mathcal{D}\lambda \exp(S),\\
\label{action1}
S&=&\int\!\! dt'\left[-\lambda \dot{Q}+{\cal H}(Q,\lambda)+(\chi/C)\Theta(t')Q\right],\,\,
\end{eqnarray}
where ${\cal H}(Q,\lambda)$ is the cumulant generating function for the current fluctuations in the quantum conductor
\begin{equation}
\nonumber
\KL I^n\KR = \partial^n_\lambda {\cal H}(Q,\Lambda)|_{\lambda = 0},
\end{equation}
and the function $\Theta(t')\equiv\theta(t')\theta(t-t')$ projects onto the interval $[0,t]$. 

Since we consider the classical noise, we are obliged to choose the leading order saddle-point solution of the SPI (\ref{SPI}), which gives
\begin{equation}
\label{action2}
\log[Z(\chi)]=S_{\rm sp}(\chi).
\end{equation}
Thus, the saddle-point action $S_{\rm sp}(\chi)$ may be considered a generator of the cumulants of the field $\phi(t)-\phi(0)$. We evaluate it up to third-order terms in $\chi$ by solving classical Hamilton's equations of motion. Namely, we split the ``Hamiltonian'' in two parts,
${\cal H}={\cal H}_0+\Delta {\cal H}$, where
\begin{equation}
 H_0=-\omega_c\lambda Q+(1/2)\KL I^2\KR
 \end{equation}
accounts for the average current and the zero-frequency noise power, and the part
\begin{equation}
\Delta H=-[\partial_Q\omega_c]\lambda Q^2+\frac{1}{2}[\partial_Q\KL I^2\KR]\lambda^2 Q
+\frac{1}{6}\KL I^3\KR\lambda^3
\label{perturbation}
\end{equation}
is to be considered as a perturbation. It contains the contribution of the third cumulant of the current in the quantum conductor, while the frist and second term represent the nonlinear and ``environmental'' cascade correction, respectively.\cite{Nagaev} 

The part ${\cal H}_0$ together 
with the source term in the action (\ref{action1}) generate the equations of motion
\begin{equation}
\dot Q=-\omega_c Q+\KL I^2\KR\lambda,\quad
\dot\lambda=\omega_c\lambda-(\chi/C)\Theta(t'),
\end{equation}
which can be easily solved with the  conditions $Q=\lambda=0$ at $t'=-\infty$ and for $t'>t$ (otherwise, $\lambda$ would diverge at infinity). The solution 
has to be substituted back to the action (\ref{action1}), eventually giving the saddle-point action (\ref{action2}). Interestingly, one can show that there is no need to account for the corrections to the equations of motion from the perturbation $\Delta {\cal H}$, since they contribute to terms starting from fourth order in $\chi$. This greatly simplifies calculations.

The final result can be presented in the following form
\begin{equation}
\log[Z(\chi)]= \chi^2 J_2(t)+\chi^3 J_3(t),
\end{equation}
where the second cumulant is given by
\begin{equation}
J_2(t)=\frac{R^2\KL I^2\KR}{2  \omega_c}\left[\omega_c t+(e^{-\omega_c t}-1)\right].
\label{J2}
\end{equation}
According to the structure of the perturbation part of the Hamiltonian (\ref{perturbation}), the third cumulant contains  three terms
\begin{equation}
J_3(t)=J_3^{\rm nl}(t)+J_3^{\rm env}(t)+J_3^{\rm min}(t)
\end{equation}
that represent the nonlinear and environmental correction, as well as the so-called minimal correlation contribution \citep{Nagaev,cascade}. Introducing the notation $\tau=\omega_ct$, they read
\begin{align}\label{J3}
&J_3^{\rm nl}=-\frac{R^3\KL I^2\KR^2\partial_Q \omega_c }{4 \omega_c^3}\left(4\tau+6\tau e^{-\tau}+e^{-2\tau} +  8e^{-\tau}-9\right),\nonumber\\ 
&J_3^{\rm env}=\frac{R^3\KL I^2\KR\partial_Q \KL I^2\KR}{2 \omega_c^2}\left[\tau(1+e^{-\tau})  +2(e^{-\tau}-1)\right],\nonumber\\
&J_3^{\rm min}=\frac{R^3\KL I^3\KR}{12 \omega_c}\left(2\tau -3+4e^{-\tau}  -e^{-2\tau}\right).
\end{align}
\normalsize
We note that the  calculations in this Appendix and the above results hold for $t>0$. For $t<0$, one can use the symmetry   $J_2(t)=J_2(-t)$ and $J_3(t)=-J_3(-t)$. Finally, evaluating the asymptotic of the expressions (\ref{J3}) for $\omega_ct\ll 1$ and $\omega_ct\gg 1$, one arrives at the results (\ref{short-time}-\ref{cascade_f}).

\bibliographystyle{apsrev4-1}
\bibliography{drag}

\begin{thebibliography}{39}%
\makeatletter
\providecommand \@ifxundefined [1]{%
 \@ifx{#1\undefined}
}%
\providecommand \@ifnum [1]{%
 \ifnum #1\expandafter \@firstoftwo
 \else \expandafter \@secondoftwo
 \fi
}%
\providecommand \@ifx [1]{%
 \ifx #1\expandafter \@firstoftwo
 \else \expandafter \@secondoftwo
 \fi
}%
\providecommand \natexlab [1]{#1}%
\providecommand \enquote  [1]{``#1''}%
\providecommand \bibnamefont  [1]{#1}%
\providecommand \bibfnamefont [1]{#1}%
\providecommand \citenamefont [1]{#1}%
\providecommand \href@noop [0]{\@secondoftwo}%
\providecommand \href [0]{\begingroup \@sanitize@url \@href}%
\providecommand \@href[1]{\@@startlink{#1}\@@href}%
\providecommand \@@href[1]{\endgroup#1\@@endlink}%
\providecommand \@sanitize@url [0]{\catcode `\\12\catcode `\$12\catcode
  `\&12\catcode `\#12\catcode `\^12\catcode `\_12\catcode `\%12\relax}%
\providecommand \@@startlink[1]{}%
\providecommand \@@endlink[0]{}%
\providecommand \url  [0]{\begingroup\@sanitize@url \@url }%
\providecommand \@url [1]{\endgroup\@href {#1}{\urlprefix }}%
\providecommand \urlprefix  [0]{URL }%
\providecommand \Eprint [0]{\href }%
\providecommand \doibase [0]{http://dx.doi.org/}%
\providecommand \selectlanguage [0]{\@gobble}%
\providecommand \bibinfo  [0]{\@secondoftwo}%
\providecommand \bibfield  [0]{\@secondoftwo}%
\providecommand \translation [1]{[#1]}%
\providecommand \BibitemOpen [0]{}%
\providecommand \bibitemStop [0]{}%
\providecommand \bibitemNoStop [0]{.\EOS\space}%
\providecommand \EOS [0]{\spacefactor3000\relax}%
\providecommand \BibitemShut  [1]{\csname bibitem#1\endcsname}%
\let\auto@bib@innerbib\@empty
\bibitem [{Note1()}]{Note1}%
  \BibitemOpen
  \bibinfo {note} {For a review, see B. N. Narozhny and A. Levchenko, Rev. Mod.
  Phys. {\protect \bf 88}, 025003 (2016), and references therein}\BibitemShut
  {NoStop}%
\bibitem [{\citenamefont {Debray}\ \emph {et~al.}(2001)\citenamefont {Debray},
  \citenamefont {Zverev}, \citenamefont {Raichev}, \citenamefont {Klesse},
  \citenamefont {Vasilopoulos},\ and\ \citenamefont {Newrock}}]{exp1d1}%
  \BibitemOpen
  \bibfield  {author} {\bibinfo {author} {\bibfnamefont {P.}~\bibnamefont
  {Debray}}, \bibinfo {author} {\bibfnamefont {V.}~\bibnamefont {Zverev}},
  \bibinfo {author} {\bibfnamefont {O.}~\bibnamefont {Raichev}}, \bibinfo
  {author} {\bibfnamefont {R.}~\bibnamefont {Klesse}}, \bibinfo {author}
  {\bibfnamefont {P.}~\bibnamefont {Vasilopoulos}}, \ and\ \bibinfo {author}
  {\bibfnamefont {R.~S.}\ \bibnamefont {Newrock}},\ }\href
  {http://stacks.iop.org/0953-8984/13/i=14/a=312} {\bibfield  {journal}
  {\bibinfo  {journal} {Journal of Physics: Condensed Matter}\ }\textbf
  {\bibinfo {volume} {13}},\ \bibinfo {pages} {3389} (\bibinfo {year}
  {2001})}\BibitemShut {NoStop}%
\bibitem [{\citenamefont {Morimoto}\ \emph {et~al.}(2003)\citenamefont
  {Morimoto}, \citenamefont {Iwase}, \citenamefont {Aoki}, \citenamefont
  {Sasaki}, \citenamefont {Ochiai}, \citenamefont {Shailos}, \citenamefont
  {Bird}, \citenamefont {Lilly}, \citenamefont {Reno},\ and\ \citenamefont
  {Simmons}}]{exp1d2}%
  \BibitemOpen
  \bibfield  {author} {\bibinfo {author} {\bibfnamefont {T.}~\bibnamefont
  {Morimoto}}, \bibinfo {author} {\bibfnamefont {Y.}~\bibnamefont {Iwase}},
  \bibinfo {author} {\bibfnamefont {N.}~\bibnamefont {Aoki}}, \bibinfo {author}
  {\bibfnamefont {T.}~\bibnamefont {Sasaki}}, \bibinfo {author} {\bibfnamefont
  {Y.}~\bibnamefont {Ochiai}}, \bibinfo {author} {\bibfnamefont
  {A.}~\bibnamefont {Shailos}}, \bibinfo {author} {\bibfnamefont {J.~P.}\
  \bibnamefont {Bird}}, \bibinfo {author} {\bibfnamefont {M.~P.}\ \bibnamefont
  {Lilly}}, \bibinfo {author} {\bibfnamefont {J.~L.}\ \bibnamefont {Reno}}, \
  and\ \bibinfo {author} {\bibfnamefont {J.~A.}\ \bibnamefont {Simmons}},\
  }\href {\doibase 10.1063/1.1579851} {\bibfield  {journal} {\bibinfo
  {journal} {Applied Physics Letters}\ }\textbf {\bibinfo {volume} {82}},\
  \bibinfo {pages} {3952} (\bibinfo {year} {2003})}\BibitemShut {NoStop}%
\bibitem [{\citenamefont {Yamamoto}\ \emph {et~al.}(2006)\citenamefont
  {Yamamoto}, \citenamefont {Stopa}, \citenamefont {Tokura}, \citenamefont
  {Hirayama},\ and\ \citenamefont {Tarucha}}]{exp1d3}%
  \BibitemOpen
  \bibfield  {author} {\bibinfo {author} {\bibfnamefont {M.}~\bibnamefont
  {Yamamoto}}, \bibinfo {author} {\bibfnamefont {M.}~\bibnamefont {Stopa}},
  \bibinfo {author} {\bibfnamefont {Y.}~\bibnamefont {Tokura}}, \bibinfo
  {author} {\bibfnamefont {Y.}~\bibnamefont {Hirayama}}, \ and\ \bibinfo
  {author} {\bibfnamefont {S.}~\bibnamefont {Tarucha}},\ }\href {\doibase
  10.1126/science.1126601} {\bibfield  {journal} {\bibinfo  {journal}
  {Science}\ }\textbf {\bibinfo {volume} {313}},\ \bibinfo {pages} {204}
  (\bibinfo {year} {2006})}\BibitemShut {NoStop}%
\bibitem [{\citenamefont {Laroche}\ \emph {et~al.}(2011)\citenamefont
  {Laroche}, \citenamefont {Gervais}, \citenamefont {Lilly},\ and\
  \citenamefont {Reno}}]{exp1d4}%
  \BibitemOpen
  \bibfield  {author} {\bibinfo {author} {\bibfnamefont {D.}~\bibnamefont
  {Laroche}}, \bibinfo {author} {\bibfnamefont {G.}~\bibnamefont {Gervais}},
  \bibinfo {author} {\bibfnamefont {M.~P.}\ \bibnamefont {Lilly}}, \ and\
  \bibinfo {author} {\bibfnamefont {J.~L.}\ \bibnamefont {Reno}},\ }\href
  {\doibase 10.1038/nnano.2011.182} {\bibfield  {journal} {\bibinfo  {journal}
  {Nature Nanotechnology}\ }\textbf {\bibinfo {volume} {6}},\ \bibinfo {pages}
  {793} (\bibinfo {year} {2011})}\BibitemShut {NoStop}%
\bibitem [{\citenamefont {Laroche}\ \emph {et~al.}(2014)\citenamefont
  {Laroche}, \citenamefont {Gervais}, \citenamefont {Lilly},\ and\
  \citenamefont {Reno}}]{exp1d5}%
  \BibitemOpen
  \bibfield  {author} {\bibinfo {author} {\bibfnamefont {D.}~\bibnamefont
  {Laroche}}, \bibinfo {author} {\bibfnamefont {G.}~\bibnamefont {Gervais}},
  \bibinfo {author} {\bibfnamefont {M.~P.}\ \bibnamefont {Lilly}}, \ and\
  \bibinfo {author} {\bibfnamefont {J.~L.}\ \bibnamefont {Reno}},\ }\href
  {\doibase 10.1126/science.1244152} {\bibfield  {journal} {\bibinfo  {journal}
  {Science}\ }\textbf {\bibinfo {volume} {343}},\ \bibinfo {pages} {631}
  (\bibinfo {year} {2014})}\BibitemShut {NoStop}%
\bibitem [{\citenamefont {Nazarov}\ and\ \citenamefont
  {Averin}(1998)}]{the1d1}%
  \BibitemOpen
  \bibfield  {author} {\bibinfo {author} {\bibfnamefont {Y.~V.}\ \bibnamefont
  {Nazarov}}\ and\ \bibinfo {author} {\bibfnamefont {D.~V.}\ \bibnamefont
  {Averin}},\ }\href {\doibase 10.1103/PhysRevLett.81.653} {\bibfield
  {journal} {\bibinfo  {journal} {Phys. Rev. Lett.}\ }\textbf {\bibinfo
  {volume} {81}},\ \bibinfo {pages} {653} (\bibinfo {year} {1998})}\BibitemShut
  {NoStop}%
\bibitem [{\citenamefont {Pustilnik}\ \emph {et~al.}(2003)\citenamefont
  {Pustilnik}, \citenamefont {Mishchenko}, \citenamefont {Glazman},\ and\
  \citenamefont {Andreev}}]{the1d2}%
  \BibitemOpen
  \bibfield  {author} {\bibinfo {author} {\bibfnamefont {M.}~\bibnamefont
  {Pustilnik}}, \bibinfo {author} {\bibfnamefont {E.~G.}\ \bibnamefont
  {Mishchenko}}, \bibinfo {author} {\bibfnamefont {L.~I.}\ \bibnamefont
  {Glazman}}, \ and\ \bibinfo {author} {\bibfnamefont {A.~V.}\ \bibnamefont
  {Andreev}},\ }\href {\doibase 10.1103/PhysRevLett.91.126805} {\bibfield
  {journal} {\bibinfo  {journal} {Phys. Rev. Lett.}\ }\textbf {\bibinfo
  {volume} {91}},\ \bibinfo {pages} {126805} (\bibinfo {year}
  {2003})}\BibitemShut {NoStop}%
\bibitem [{\citenamefont {Onac}\ \emph {et~al.}(2006)\citenamefont {Onac},
  \citenamefont {Balestro}, \citenamefont {van Beveren}, \citenamefont
  {Hartmann}, \citenamefont {Nazarov},\ and\ \citenamefont
  {Kouwenhoven}}]{expQD1}%
  \BibitemOpen
  \bibfield  {author} {\bibinfo {author} {\bibfnamefont {E.}~\bibnamefont
  {Onac}}, \bibinfo {author} {\bibfnamefont {F.}~\bibnamefont {Balestro}},
  \bibinfo {author} {\bibfnamefont {L.~H.~W.}\ \bibnamefont {van Beveren}},
  \bibinfo {author} {\bibfnamefont {U.}~\bibnamefont {Hartmann}}, \bibinfo
  {author} {\bibfnamefont {Y.~V.}\ \bibnamefont {Nazarov}}, \ and\ \bibinfo
  {author} {\bibfnamefont {L.~P.}\ \bibnamefont {Kouwenhoven}},\ }\href
  {\doibase 10.1103/PhysRevLett.96.176601} {\bibfield  {journal} {\bibinfo
  {journal} {Phys. Rev. Lett.}\ }\textbf {\bibinfo {volume} {96}},\ \bibinfo
  {pages} {176601} (\bibinfo {year} {2006})}\BibitemShut {NoStop}%
\bibitem [{\citenamefont {Aguado}\ and\ \citenamefont
  {Kouwenhoven}(2000)}]{theQD1}%
  \BibitemOpen
  \bibfield  {author} {\bibinfo {author} {\bibfnamefont {R.}~\bibnamefont
  {Aguado}}\ and\ \bibinfo {author} {\bibfnamefont {L.~P.}\ \bibnamefont
  {Kouwenhoven}},\ }\href {\doibase 10.1103/PhysRevLett.84.1986} {\bibfield
  {journal} {\bibinfo  {journal} {Phys. Rev. Lett.}\ }\textbf {\bibinfo
  {volume} {84}},\ \bibinfo {pages} {1986} (\bibinfo {year}
  {2000})}\BibitemShut {NoStop}%
\bibitem [{\citenamefont {S\'anchez}\ \emph {et~al.}(2010)\citenamefont
  {S\'anchez}, \citenamefont {L\'opez}, \citenamefont {S\'anchez},\ and\
  \citenamefont {B\"uttiker}}]{theQD2}%
  \BibitemOpen
  \bibfield  {author} {\bibinfo {author} {\bibfnamefont {R.}~\bibnamefont
  {S\'anchez}}, \bibinfo {author} {\bibfnamefont {R.}~\bibnamefont {L\'opez}},
  \bibinfo {author} {\bibfnamefont {D.}~\bibnamefont {S\'anchez}}, \ and\
  \bibinfo {author} {\bibfnamefont {M.}~\bibnamefont {B\"uttiker}},\ }\href
  {\doibase 10.1103/PhysRevLett.104.076801} {\bibfield  {journal} {\bibinfo
  {journal} {Phys. Rev. Lett.}\ }\textbf {\bibinfo {volume} {104}},\ \bibinfo
  {pages} {076801} (\bibinfo {year} {2010})}\BibitemShut {NoStop}%
\bibitem [{\citenamefont {Sothmann}\ \emph {et~al.}(2012)\citenamefont
  {Sothmann}, \citenamefont {S\'anchez}, \citenamefont {Jordan},\ and\
  \citenamefont {B\"uttiker}}]{theQD3}%
  \BibitemOpen
  \bibfield  {author} {\bibinfo {author} {\bibfnamefont {B.}~\bibnamefont
  {Sothmann}}, \bibinfo {author} {\bibfnamefont {R.}~\bibnamefont {S\'anchez}},
  \bibinfo {author} {\bibfnamefont {A.~N.}\ \bibnamefont {Jordan}}, \ and\
  \bibinfo {author} {\bibfnamefont {M.}~\bibnamefont {B\"uttiker}},\ }\href
  {\doibase 10.1103/PhysRevB.85.205301} {\bibfield  {journal} {\bibinfo
  {journal} {Phys. Rev. B}\ }\textbf {\bibinfo {volume} {85}},\ \bibinfo
  {pages} {205301} (\bibinfo {year} {2012})}\BibitemShut {NoStop}%
\bibitem [{\citenamefont {Khrapai}\ \emph {et~al.}(2006)\citenamefont
  {Khrapai}, \citenamefont {Ludwig}, \citenamefont {Kotthaus}, \citenamefont
  {Tranitz},\ and\ \citenamefont {Wegscheider}}]{expqpc1}%
  \BibitemOpen
  \bibfield  {author} {\bibinfo {author} {\bibfnamefont {V.~S.}\ \bibnamefont
  {Khrapai}}, \bibinfo {author} {\bibfnamefont {S.}~\bibnamefont {Ludwig}},
  \bibinfo {author} {\bibfnamefont {J.~P.}\ \bibnamefont {Kotthaus}}, \bibinfo
  {author} {\bibfnamefont {H.~P.}\ \bibnamefont {Tranitz}}, \ and\ \bibinfo
  {author} {\bibfnamefont {W.}~\bibnamefont {Wegscheider}},\ }\href {\doibase
  10.1103/PhysRevLett.97.176803} {\bibfield  {journal} {\bibinfo  {journal}
  {Phys. Rev. Lett.}\ }\textbf {\bibinfo {volume} {97}},\ \bibinfo {pages}
  {176803} (\bibinfo {year} {2006})}\BibitemShut {NoStop}%
\bibitem [{\citenamefont {Khrapai}\ \emph {et~al.}(2007)\citenamefont
  {Khrapai}, \citenamefont {Ludwig}, \citenamefont {Kotthaus}, \citenamefont
  {Tranitz},\ and\ \citenamefont {Wegscheider}}]{expqpc2}%
  \BibitemOpen
  \bibfield  {author} {\bibinfo {author} {\bibfnamefont {V.~S.}\ \bibnamefont
  {Khrapai}}, \bibinfo {author} {\bibfnamefont {S.}~\bibnamefont {Ludwig}},
  \bibinfo {author} {\bibfnamefont {J.~P.}\ \bibnamefont {Kotthaus}}, \bibinfo
  {author} {\bibfnamefont {H.~P.}\ \bibnamefont {Tranitz}}, \ and\ \bibinfo
  {author} {\bibfnamefont {W.}~\bibnamefont {Wegscheider}},\ }\href {\doibase
  10.1103/PhysRevLett.99.096803} {\bibfield  {journal} {\bibinfo  {journal}
  {Phys. Rev. Lett.}\ }\textbf {\bibinfo {volume} {99}},\ \bibinfo {pages}
  {096803} (\bibinfo {year} {2007})}\BibitemShut {NoStop}%
\bibitem [{\citenamefont {Levchenko}\ and\ \citenamefont
  {Kamenev}(2008{\natexlab{a}})}]{kamenev2}%
  \BibitemOpen
  \bibfield  {author} {\bibinfo {author} {\bibfnamefont {A.}~\bibnamefont
  {Levchenko}}\ and\ \bibinfo {author} {\bibfnamefont {A.}~\bibnamefont
  {Kamenev}},\ }\href {\doibase 10.1103/PhysRevLett.101.216806} {\bibfield
  {journal} {\bibinfo  {journal} {Phys. Rev. Lett.}\ }\textbf {\bibinfo
  {volume} {101}},\ \bibinfo {pages} {216806} (\bibinfo {year}
  {2008}{\natexlab{a}})}\BibitemShut {NoStop}%
\bibitem [{\citenamefont {Blanter}\ and\ \citenamefont
  {Büttiker}(2000)}]{Blanter-Buttiker}%
  \BibitemOpen
  \bibfield  {author} {\bibinfo {author} {\bibfnamefont {Y.}~\bibnamefont
  {Blanter}}\ and\ \bibinfo {author} {\bibfnamefont {M.}~\bibnamefont
  {Büttiker}},\ }\href {\doibase
  https://doi.org/10.1016/S0370-1573(99)00123-4} {\bibfield  {journal}
  {\bibinfo  {journal} {Physics Reports}\ }\textbf {\bibinfo {volume} {336}},\
  \bibinfo {pages} {1 } (\bibinfo {year} {2000})}\BibitemShut {NoStop}%
\bibitem [{\citenamefont {Kamenev}\ and\ \citenamefont
  {Oreg}(1995)}]{kamenev1}%
  \BibitemOpen
  \bibfield  {author} {\bibinfo {author} {\bibfnamefont {A.}~\bibnamefont
  {Kamenev}}\ and\ \bibinfo {author} {\bibfnamefont {Y.}~\bibnamefont {Oreg}},\
  }\href {\doibase 10.1103/PhysRevB.52.7516} {\bibfield  {journal} {\bibinfo
  {journal} {Phys. Rev. B}\ }\textbf {\bibinfo {volume} {52}},\ \bibinfo
  {pages} {7516} (\bibinfo {year} {1995})}\BibitemShut {NoStop}%
\bibitem [{\citenamefont {Levitov}\ \emph {et~al.}(1996)\citenamefont
  {Levitov}, \citenamefont {Lee},\ and\ \citenamefont {Lesovik}}]{Levitov}%
  \BibitemOpen
  \bibfield  {author} {\bibinfo {author} {\bibfnamefont {L.~S.}\ \bibnamefont
  {Levitov}}, \bibinfo {author} {\bibfnamefont {H.}~\bibnamefont {Lee}}, \ and\
  \bibinfo {author} {\bibfnamefont {G.~B.}\ \bibnamefont {Lesovik}},\ }\href
  {\doibase 10.1063/1.531672} {\bibfield  {journal} {\bibinfo  {journal}
  {Journal of Mathematical Physics}\ }\textbf {\bibinfo {volume} {37}},\
  \bibinfo {pages} {4845} (\bibinfo {year} {1996})}\BibitemShut {NoStop}%
\bibitem [{\citenamefont {Reulet}\ \emph {et~al.}(2003)\citenamefont {Reulet},
  \citenamefont {Senzier},\ and\ \citenamefont {Prober}}]{3dmom1}%
  \BibitemOpen
  \bibfield  {author} {\bibinfo {author} {\bibfnamefont {B.}~\bibnamefont
  {Reulet}}, \bibinfo {author} {\bibfnamefont {J.}~\bibnamefont {Senzier}}, \
  and\ \bibinfo {author} {\bibfnamefont {D.~E.}\ \bibnamefont {Prober}},\
  }\href {\doibase 10.1103/PhysRevLett.91.196601} {\bibfield  {journal}
  {\bibinfo  {journal} {Phys. Rev. Lett.}\ }\textbf {\bibinfo {volume} {91}},\
  \bibinfo {pages} {196601} (\bibinfo {year} {2003})}\BibitemShut {NoStop}%
\bibitem [{\citenamefont {Bomze}\ \emph {et~al.}(2005)\citenamefont {Bomze},
  \citenamefont {Gershon}, \citenamefont {Shovkun}, \citenamefont {Levitov},\
  and\ \citenamefont {Reznikov}}]{3dmom2}%
  \BibitemOpen
  \bibfield  {author} {\bibinfo {author} {\bibfnamefont {Y.}~\bibnamefont
  {Bomze}}, \bibinfo {author} {\bibfnamefont {G.}~\bibnamefont {Gershon}},
  \bibinfo {author} {\bibfnamefont {D.}~\bibnamefont {Shovkun}}, \bibinfo
  {author} {\bibfnamefont {L.~S.}\ \bibnamefont {Levitov}}, \ and\ \bibinfo
  {author} {\bibfnamefont {M.}~\bibnamefont {Reznikov}},\ }\href {\doibase
  10.1103/PhysRevLett.95.176601} {\bibfield  {journal} {\bibinfo  {journal}
  {Phys. Rev. Lett.}\ }\textbf {\bibinfo {volume} {95}},\ \bibinfo {pages}
  {176601} (\bibinfo {year} {2005})}\BibitemShut {NoStop}%
\bibitem [{\citenamefont {Gustavsson}\ \emph {et~al.}(2006)\citenamefont
  {Gustavsson}, \citenamefont {Leturcq}, \citenamefont
  {Simovi\ifmmode~\check{c}\else \v{c}\fi{}}, \citenamefont {Schleser},
  \citenamefont {Ihn}, \citenamefont {Studerus}, \citenamefont {Ensslin},
  \citenamefont {Driscoll},\ and\ \citenamefont {Gossard}}]{3dmom3}%
  \BibitemOpen
  \bibfield  {author} {\bibinfo {author} {\bibfnamefont {S.}~\bibnamefont
  {Gustavsson}}, \bibinfo {author} {\bibfnamefont {R.}~\bibnamefont {Leturcq}},
  \bibinfo {author} {\bibfnamefont {B.}~\bibnamefont
  {Simovi\ifmmode~\check{c}\else \v{c}\fi{}}}, \bibinfo {author} {\bibfnamefont
  {R.}~\bibnamefont {Schleser}}, \bibinfo {author} {\bibfnamefont
  {T.}~\bibnamefont {Ihn}}, \bibinfo {author} {\bibfnamefont {P.}~\bibnamefont
  {Studerus}}, \bibinfo {author} {\bibfnamefont {K.}~\bibnamefont {Ensslin}},
  \bibinfo {author} {\bibfnamefont {D.~C.}\ \bibnamefont {Driscoll}}, \ and\
  \bibinfo {author} {\bibfnamefont {A.~C.}\ \bibnamefont {Gossard}},\ }\href
  {\doibase 10.1103/PhysRevLett.96.076605} {\bibfield  {journal} {\bibinfo
  {journal} {Phys. Rev. Lett.}\ }\textbf {\bibinfo {volume} {96}},\ \bibinfo
  {pages} {076605} (\bibinfo {year} {2006})}\BibitemShut {NoStop}%
\bibitem [{\citenamefont {Gershon}\ \emph {et~al.}(2008)\citenamefont
  {Gershon}, \citenamefont {Bomze}, \citenamefont {Sukhorukov},\ and\
  \citenamefont {Reznikov}}]{3dmom4}%
  \BibitemOpen
  \bibfield  {author} {\bibinfo {author} {\bibfnamefont {G.}~\bibnamefont
  {Gershon}}, \bibinfo {author} {\bibfnamefont {Y.}~\bibnamefont {Bomze}},
  \bibinfo {author} {\bibfnamefont {E.~V.}\ \bibnamefont {Sukhorukov}}, \ and\
  \bibinfo {author} {\bibfnamefont {M.}~\bibnamefont {Reznikov}},\ }\href
  {\doibase 10.1103/PhysRevLett.101.016803} {\bibfield  {journal} {\bibinfo
  {journal} {Phys. Rev. Lett.}\ }\textbf {\bibinfo {volume} {101}},\ \bibinfo
  {pages} {016803} (\bibinfo {year} {2008})}\BibitemShut {NoStop}%
\bibitem [{\citenamefont {Gustavsson}\ \emph {et~al.}(2009)\citenamefont
  {Gustavsson}, \citenamefont {Leturcq}, \citenamefont {Studer}, \citenamefont
  {Shorubalko}, \citenamefont {Ihn}, \citenamefont {Ensslin}, \citenamefont
  {Driscoll},\ and\ \citenamefont {Gossard}}]{3dmom5}%
  \BibitemOpen
  \bibfield  {author} {\bibinfo {author} {\bibfnamefont {S.}~\bibnamefont
  {Gustavsson}}, \bibinfo {author} {\bibfnamefont {R.}~\bibnamefont {Leturcq}},
  \bibinfo {author} {\bibfnamefont {M.}~\bibnamefont {Studer}}, \bibinfo
  {author} {\bibfnamefont {I.}~\bibnamefont {Shorubalko}}, \bibinfo {author}
  {\bibfnamefont {T.}~\bibnamefont {Ihn}}, \bibinfo {author} {\bibfnamefont
  {K.}~\bibnamefont {Ensslin}}, \bibinfo {author} {\bibfnamefont
  {D.}~\bibnamefont {Driscoll}}, \ and\ \bibinfo {author} {\bibfnamefont
  {A.}~\bibnamefont {Gossard}},\ }\href {\doibase
  https://doi.org/10.1016/j.surfrep.2009.02.001} {\bibfield  {journal}
  {\bibinfo  {journal} {Surface Science Reports}\ }\textbf {\bibinfo {volume}
  {64}},\ \bibinfo {pages} {191 } (\bibinfo {year} {2009})}\BibitemShut
  {NoStop}%
\bibitem [{\citenamefont {Gabelli}\ and\ \citenamefont
  {Reulet}(2009)}]{3dmom6}%
  \BibitemOpen
  \bibfield  {author} {\bibinfo {author} {\bibfnamefont {J.}~\bibnamefont
  {Gabelli}}\ and\ \bibinfo {author} {\bibfnamefont {B.}~\bibnamefont
  {Reulet}},\ }\href {\doibase 10.1103/PhysRevB.80.161203} {\bibfield
  {journal} {\bibinfo  {journal} {Phys. Rev. B}\ }\textbf {\bibinfo {volume}
  {80}},\ \bibinfo {pages} {161203} (\bibinfo {year} {2009})}\BibitemShut
  {NoStop}%
\bibitem [{\citenamefont {Timofeev}\ \emph {et~al.}(2007)\citenamefont
  {Timofeev}, \citenamefont {Meschke}, \citenamefont {Peltonen}, \citenamefont
  {Heikkil\"a},\ and\ \citenamefont {Pekola}}]{3dmom7}%
  \BibitemOpen
  \bibfield  {author} {\bibinfo {author} {\bibfnamefont {A.~V.}\ \bibnamefont
  {Timofeev}}, \bibinfo {author} {\bibfnamefont {M.}~\bibnamefont {Meschke}},
  \bibinfo {author} {\bibfnamefont {J.~T.}\ \bibnamefont {Peltonen}}, \bibinfo
  {author} {\bibfnamefont {T.~T.}\ \bibnamefont {Heikkil\"a}}, \ and\ \bibinfo
  {author} {\bibfnamefont {J.~P.}\ \bibnamefont {Pekola}},\ }\href {\doibase
  10.1103/PhysRevLett.98.207001} {\bibfield  {journal} {\bibinfo  {journal}
  {Phys. Rev. Lett.}\ }\textbf {\bibinfo {volume} {98}},\ \bibinfo {pages}
  {207001} (\bibinfo {year} {2007})}\BibitemShut {NoStop}%
\bibitem [{\citenamefont {Huard}\ \emph {et~al.}(2007)\citenamefont {Huard},
  \citenamefont {Pothier}, \citenamefont {Birge}, \citenamefont {Est\`eve},
  \citenamefont {Waintal},\ and\ \citenamefont {Ankerhold}}]{3dmom8}%
  \BibitemOpen
  \bibfield  {author} {\bibinfo {author} {\bibfnamefont {B.}~\bibnamefont
  {Huard}}, \bibinfo {author} {\bibfnamefont {H.}~\bibnamefont {Pothier}},
  \bibinfo {author} {\bibfnamefont {N.}~\bibnamefont {Birge}}, \bibinfo
  {author} {\bibfnamefont {D.}~\bibnamefont {Est\`eve}}, \bibinfo {author}
  {\bibfnamefont {X.}~\bibnamefont {Waintal}}, \ and\ \bibinfo {author}
  {\bibfnamefont {J.}~\bibnamefont {Ankerhold}},\ }\href {\doibase
  10.1002/andp.200710263} {\bibfield  {journal} {\bibinfo  {journal} {Ann.
  Phys. (Leipzig)}\ }\textbf {\bibinfo {volume} {16}},\ \bibinfo {pages} {736}
  (\bibinfo {year} {2007})}\BibitemShut {NoStop}%
\bibitem [{\citenamefont {Sukhorukov}\ and\ \citenamefont
  {Edwards}(2008)}]{edwards}%
  \BibitemOpen
  \bibfield  {author} {\bibinfo {author} {\bibfnamefont {E.~V.}\ \bibnamefont
  {Sukhorukov}}\ and\ \bibinfo {author} {\bibfnamefont {J.}~\bibnamefont
  {Edwards}},\ }\href {\doibase 10.1103/PhysRevB.78.035332} {\bibfield
  {journal} {\bibinfo  {journal} {Phys. Rev. B}\ }\textbf {\bibinfo {volume}
  {78}},\ \bibinfo {pages} {035332} (\bibinfo {year} {2008})}\BibitemShut
  {NoStop}%
\bibitem [{\citenamefont {Levchenko}\ and\ \citenamefont
  {Kamenev}(2008{\natexlab{b}})}]{levch}%
  \BibitemOpen
  \bibfield  {author} {\bibinfo {author} {\bibfnamefont {A.}~\bibnamefont
  {Levchenko}}\ and\ \bibinfo {author} {\bibfnamefont {A.}~\bibnamefont
  {Kamenev}},\ }\href {\doibase 10.1103/PhysRevLett.100.026805} {\bibfield
  {journal} {\bibinfo  {journal} {Phys. Rev. Lett.}\ }\textbf {\bibinfo
  {volume} {100}},\ \bibinfo {pages} {026805} (\bibinfo {year}
  {2008}{\natexlab{b}})}\BibitemShut {NoStop}%
\bibitem [{Note2()}]{Note2}%
  \BibitemOpen
  \bibinfo {note} {The quantum drag effect to third order in coupling has been
  studied in [\protect \rev@citealpnum {levch}]}\BibitemShut {NoStop}%
\bibitem [{\citenamefont {Geigenm\"uller}\ and\ \citenamefont
  {Nazarov}(1991)}]{Nazarov1}%
  \BibitemOpen
  \bibfield  {author} {\bibinfo {author} {\bibfnamefont {U.}~\bibnamefont
  {Geigenm\"uller}}\ and\ \bibinfo {author} {\bibfnamefont {Y.~V.}\
  \bibnamefont {Nazarov}},\ }\href {\doibase 10.1103/PhysRevB.44.10953}
  {\bibfield  {journal} {\bibinfo  {journal} {Phys. Rev. B}\ }\textbf {\bibinfo
  {volume} {44}},\ \bibinfo {pages} {10953} (\bibinfo {year}
  {1991})}\BibitemShut {NoStop}%
\bibitem [{\citenamefont {Ingold}\ and\ \citenamefont {Nazarov}(1992)}]{Naz}%
  \BibitemOpen
  \bibfield  {author} {\bibinfo {author} {\bibfnamefont {G.}~\bibnamefont
  {Ingold}}\ and\ \bibinfo {author} {\bibfnamefont {Y.}~\bibnamefont
  {Nazarov}},\ }in\ \href@noop {} {\emph {\bibinfo {booktitle} {Single Charge
  Tunneling}}},\ \bibinfo {series} {NATO ASI Series(Series B: Physics)}, Vol.\
  \bibinfo {volume} {294},\ \bibinfo {editor} {edited by\ \bibinfo {editor}
  {\bibfnamefont {H.}~\bibnamefont {Grabert}}\ and\ \bibinfo {editor}
  {\bibfnamefont {M.}~\bibnamefont {Devoret}}}\ (\bibinfo  {publisher}
  {Springer},\ \bibinfo {address} {Boston, MA},\ \bibinfo {year}
  {1992})\BibitemShut {NoStop}%
\bibitem [{\citenamefont {Pilgram}\ \emph {et~al.}(2003)\citenamefont
  {Pilgram}, \citenamefont {Jordan}, \citenamefont {Sukhorukov},\ and\
  \citenamefont {B\"uttiker}}]{stochpath}%
  \BibitemOpen
  \bibfield  {author} {\bibinfo {author} {\bibfnamefont {S.}~\bibnamefont
  {Pilgram}}, \bibinfo {author} {\bibfnamefont {A.~N.}\ \bibnamefont {Jordan}},
  \bibinfo {author} {\bibfnamefont {E.~V.}\ \bibnamefont {Sukhorukov}}, \ and\
  \bibinfo {author} {\bibfnamefont {M.}~\bibnamefont {B\"uttiker}},\ }\href
  {\doibase 10.1103/PhysRevLett.90.206801} {\bibfield  {journal} {\bibinfo
  {journal} {Phys. Rev. Lett.}\ }\textbf {\bibinfo {volume} {90}},\ \bibinfo
  {pages} {206801} (\bibinfo {year} {2003})}\BibitemShut {NoStop}%
\bibitem [{\citenamefont {Jordan}\ \emph {et~al.}(2004)\citenamefont {Jordan},
  \citenamefont {Sukhorukov},\ and\ \citenamefont {Pilgram}}]{stochpath2}%
  \BibitemOpen
  \bibfield  {author} {\bibinfo {author} {\bibfnamefont {A.~N.}\ \bibnamefont
  {Jordan}}, \bibinfo {author} {\bibfnamefont {E.~V.}\ \bibnamefont
  {Sukhorukov}}, \ and\ \bibinfo {author} {\bibfnamefont {S.}~\bibnamefont
  {Pilgram}},\ }\href {\doibase 10.1063/1.1803927} {\bibfield  {journal}
  {\bibinfo  {journal} {Journal of Mathematical Physics}\ }\textbf {\bibinfo
  {volume} {45}},\ \bibinfo {pages} {4386} (\bibinfo {year}
  {2004})}\BibitemShut {NoStop}%
\bibitem [{\citenamefont {{Safi}}(2014)}]{Safi1}%
  \BibitemOpen
  \bibfield  {author} {\bibinfo {author} {\bibfnamefont {I.}~\bibnamefont
  {{Safi}}},\ }\href@noop {} {\bibfield  {journal} {\bibinfo  {journal} {ArXiv
  e-prints}\ ,\ \bibinfo {eid} {arXiv:1401.5950}} (\bibinfo {year} {2014})},\
  \Eprint {http://arxiv.org/abs/1401.5950} {arXiv:1401.5950
  [cond-mat.mes-hall]} \BibitemShut {NoStop}%
\bibitem [{\citenamefont {Safi}(2019)}]{Safi2}%
  \BibitemOpen
  \bibfield  {author} {\bibinfo {author} {\bibfnamefont {I.}~\bibnamefont
  {Safi}},\ }\href {\doibase 10.1103/PhysRevB.99.045101} {\bibfield  {journal}
  {\bibinfo  {journal} {Phys. Rev. B}\ }\textbf {\bibinfo {volume} {99}},\
  \bibinfo {pages} {045101} (\bibinfo {year} {2019})}\BibitemShut {NoStop}%
\bibitem [{\citenamefont {Mahan}(1980)}]{Mahan}%
  \BibitemOpen
  \bibfield  {author} {\bibinfo {author} {\bibfnamefont {G.~D.}\ \bibnamefont
  {Mahan}},\ }\href@noop {} {\emph {\bibinfo {title} {Many-Particle Physics}}}\
  (\bibinfo  {publisher} {Plenum},\ \bibinfo {address} {New York},\ \bibinfo
  {year} {1980})\BibitemShut {NoStop}%
\bibitem [{Note3()}]{Note3}%
  \BibitemOpen
  \bibinfo {note} {Note that $J_3$ is smaller than $J_2$ by the dimensionless
  coupling constant $R\ll 1$ (in unites where $|e|=\hbar =1$), therefore one
  should consider the quantum correction to $J_2$. However, it has been shown
  in Ref.\ [\protect \rev@citealpnum {edwards}], that according to the Kubo
  linear response formula it is simply given by the differential conductance of
  the noise source, which is an equilibrium property. Therefore, it does not
  contribute to the drag effect.}\BibitemShut {Stop}%
\bibitem [{\citenamefont {Nagaev}(2002)}]{Nagaev}%
  \BibitemOpen
  \bibfield  {author} {\bibinfo {author} {\bibfnamefont {K.~E.}\ \bibnamefont
  {Nagaev}},\ }\href {\doibase 10.1103/PhysRevB.66.075334} {\bibfield
  {journal} {\bibinfo  {journal} {Phys. Rev. B}\ }\textbf {\bibinfo {volume}
  {66}},\ \bibinfo {pages} {075334} (\bibinfo {year} {2002})}\BibitemShut
  {NoStop}%
\bibitem [{\citenamefont {Beenakker}\ \emph {et~al.}(2003)\citenamefont
  {Beenakker}, \citenamefont {Kindermann},\ and\ \citenamefont
  {Nazarov}}]{cascade}%
  \BibitemOpen
  \bibfield  {author} {\bibinfo {author} {\bibfnamefont {C.~W.~J.}\
  \bibnamefont {Beenakker}}, \bibinfo {author} {\bibfnamefont {M.}~\bibnamefont
  {Kindermann}}, \ and\ \bibinfo {author} {\bibfnamefont {Y.~V.}\ \bibnamefont
  {Nazarov}},\ }\href {\doibase 10.1103/PhysRevLett.90.176802} {\bibfield
  {journal} {\bibinfo  {journal} {Phys. Rev. Lett.}\ }\textbf {\bibinfo
  {volume} {90}},\ \bibinfo {pages} {176802} (\bibinfo {year}
  {2003})}\BibitemShut {NoStop}%
\end{thebibliography}%

\end{document}